\documentclass[reqno,
]{amsproc}
\usepackage{
amssymb}
\allowdisplaybreaks[4]
\DeclareFontFamily{OT1}{wncyi}{}
\DeclareFontShape{OT1}{wncyi}{m}{it}{
   <5> <6> <7> <8> <9> gen * wncyi
   <10> <10.95> <12> <14.4> <17.28> <20.74> <24.88> wncyi10
  }{}
\DeclareSymbolFont{cyrletters}{OT1}{wncyi}{m}{it}
\DeclareSymbolFontAlphabet{\cyrmath}{cyrletters}
\DeclareMathSymbol{\re}{\cyrmath}{cyrletters}{"03}

\newlength{\SomeLenght}
\newcommand*{\hdotsas}[1]{\settowidth{\SomeLenght}{$\displaystyle#1$}
                          \hbox to\SomeLenght{\dotfill}}

\newcommand{\cprime}{\/{\mathsurround=0pt$'$}}
\newcommand{\pinner}{\mathbin{\mathchoice
   {\hbox{\vrule width0.6em depth0pt height0.4pt
   \vrule width0.4pt depth0pt height0.8ex}}
   {\hbox{\vrule width0.6em depth0pt height0.4pt
   \vrule width0.4pt depth0pt height0.8ex}}
   {\hbox{\kern0.14em
   \vrule width0.48em depth0pt height0.4pt
   \vrule width0.4pt depth0pt height0.6ex\kern0.14em}}
   {\hbox{\kern0.1em
   \vrule width0.39em depth0pt height0.4pt
   \vrule width0.4pt depth0pt height0.5ex\kern0.1em}}}}

\newcommand{\BBR}{\mathbb{R}}

\newcommand{\CR}{\mathcal{R}}

\newcommand{\om}{\omega}

\newcommand{\oth} {\overline{\theta}}

\newcommand{\oD} {\overline{D}}
\newcommand{\oQ} {\overline{Q}}
\newcommand{\wD} {\widetilde{D}}
\newcommand{\of}{{\overline{f}}}

\theoremstyle{definition}

\theoremstyle{remark}

\begin{document}
\title[Deformation for the Supersymmetric KdV]{Deformation and Recursion for the $N=2$\ 
  $\alpha=1$ Supersymmetric KdV-hierarchy}
\author{ Alexander S. Sorin}
\address{Bogoliubov Laboratory of Theoretical Physics \\
Joint Institute for Nuclear Research (JINR)\\
 141980 Dubna, Moscow Region\\
 Russia}
\email{sorin@thsun1.jinr.ru}
\author{Paul H. M. Kersten}
\address{University of Twente,
Faculty of Mathematical Sciences,
P.O.Box 217,
7500 AE Enschede,
The Netherlands}
\email{kersten@math.utwente.nl}
\keywords{Complete Integrability, Deformations, Bi-Hamiltonian Structure,
Recursion Operators, Symmetries, Conservation Laws, Coverings}
\subjclass{58F07, 58G37, 58H15, 58F37}
\begin{abstract}
A detailed description is given for the construction of the deformation
of the $N=2$ supersymmetric 
  $\alpha=1$ KdV-equation, leading to the recursion operator for
  symmetries and the zero-th Hamiltonian structure; the solution to  a
longstanding problem.
 
\end{abstract}
\maketitle

\section{\bf Introduction}
The $N=2$ supersymmetric ${\alpha}=1$ KdV-equation was originally 
introduced in \cite{lm} as a Hamiltonian equation with the $N=2$
superconformal algebra as a second Hamiltonian structure, and 
its integrability was conjectured there due to the existence of 
a few additional nontrivial bosonic Hamiltonians. Then its Lax--pair
representation has indeed been constructed in \cite{pop1}, and it 
allowed an algoritmic reconstruction of the whole tower of highest 
{\it commutative} bosonic flows and their Hamiltonians belonging to the
$N=2$ supersymmetric ${\alpha}=1$ KdV-hierarchy. 

Actually, besides the $N=2$  ${\alpha}=1$ KdV-equation
there are another two inequivalent $N=2$  supersymmetric 
Hamiltonian equations with the $N=2$  superconformal algebra as a
second Hamiltonian structure (the $N=2$  ${\alpha}=-2$ and
${\alpha}=4$ KdV-equations \cite{mat,lm}), but the $N=2$  ${\alpha}=1$
KdV-equation is rather exceptional \cite{bks}. Despite knowledge of its
Lax--pair description, there remains a lot of longstanding, unsolved
problems which resolution would be quite important for a deeper
understanding and more detailed description of the $N=2$  ${\alpha}=1$ KdV
hierarchy. Thus,
since the time when the $N=2$  ${\alpha}=1$ KdV-equation was proposed, much
efforts were made to construct a tower of its {\it noncommutative} bosonic
and fermionic, local and nonlocal symmetries and Hamiltonians,
bi-Hamiltonian structure as well as recursion operator (see, e.g.
discussions in \cite{dm,m} and references therein). Though these
rather complicated problems, solved for the case of
the $N=2$ ${\alpha}=-2$ and ${\alpha}=4$ KdV-hierarchies,
still wait their complete resolution for the $N=2$ ${\alpha}=1$ KdV-
hierarchy, a considerable progress towards their solution arose quite
recently. Thus, the {\it puzzle} \cite{dm,m}, related to the 
{\it "nonexistence"} 
of higher fermionic flows of the $N=2$ ${\alpha}=1$ KdV-hierarchy,
was partly resolved in \cite{KrasKers-book,ker1} by explicit constructing
a few
bosonic and fermionic {\it nonlocal} and {\it nonpolynomial} flows
and Hamiltonains, then their $N=2$ superfield structure and origin were
uncovered in \cite{ks}. A new property, crucial for the existence of these
flows and Hamiltonians, making them distinguished compared to all
flows and Hamiltonians of other supersymmetric hierarchies 
constructed before, is their {\it nonpolynomiality}.
A new approach to a recursion operator treating it as a form--valued
vector field which satisfies a generalized symmetry equation related to
a given equation was developed in \cite{kr1, kerkr2}. Using this approach
the recursion operator of the bosonic limit of the N=2
${\alpha}=1$ KdV-hierarchy was derived in \cite{ker2}, and its structure,
underlining relevance of these Hamiltonians in the bosonic limit,  
gives a hint towards its supersymmetric generalization.\\
The organisation of this paper is as follows.\\
First the general notions from the mathematical  theory  of symmetries,
nonlocalities,
deformations and form-valued vector fields are  exposed to some detail in
Section 2  and its subsections for the classical KdV-equation. For
full details the reader is referred to e.g. \cite{kr1,KrasKers-book}.\\
In Section 3 we expose all results obtained for the $N=2$ $ \alpha=1$
supersymmetric KdV-equation in great detail.\\
In Section 4 we present conclusions, while in Section 5, arranged as
an Appendix, results of the Poisson bracket structure are given.

\section{Nonlocal Setting for Differential Equations, the KdV-equation}

\subsection {Nonlocalities}
As standard example, to illustrate the notions we are going to
discuss for the $N=2$ $\alpha=1$ supersymmetric KdV-equation in
Section 3, we take the KdV-equation
\begin{align}\label{sec1:kdv}
u_t& = uu_x + u_{xxx}.
\end{align}
For a short  theoretical introduction we refer to \cite{ker2}, while for
more detailed expositions we refer to \cite{kr1,KrasKers-book,kerkr2}.

We consider $Y \subset J^{\infty}(x,t; u)$ the infinite prolongation
of \eqref{sec1:kdv}, c.f.\cite{KLV,KV-trends}, where coordinates in the
infinite jet bundle $J^{\infty}(x, t; u)$ are
given by $(x, t, u, u_x, u_t, \cdots )$ and $Y$ is formally described
as the submanifold of $J^{\infty}(x,t; u)$ defined by
\begin{align}\label{sec1:eq:kdv}
u_t &= uu_x + u_{xxx},\nonumber   \\
u_{xt} &= uu_{xx} + u^2_x + u_{xxxx},\\
\vdots  \quad .& \nonumber                  
\end{align}
As internal coordinates in $Y$ one chooses $(x, t, u, u_x, u_{xx},
\cdots )$ while $u_t, u_{xt}, \cdots$ are obtained from \eqref{sec1:eq:kdv}.\\

The Cartan distribution on $Y$ is given by the total partial derivative
vector fields
\begin{align}
\begin{array}{lclcl}
\wD_x &=& \partial_x &+& \sum\limits_{n\geq 0} u_{n+1}\partial_{u_n},\\
\wD_t &=& \partial_t &+& \sum\limits_{n\geq 0} u_{nt} \partial_{u_n}
\end{array}
\end{align}
where $u=u_0,~u_1 = u_x, ~u_2 = u_{xx} , \cdots \quad ; ~u_{1t} = u_{xt};
~u_{2t} = u_{xxt} \cdots$.\\

Classically the notion of a \emph{generalized} or \emph{higher symmetry} 
$Y$ of a differential equation $\{F=0\}$ is defined as a vertical
vector field V
\begin{equation}\label{sec1:eq:vec}
        V = \re_f = f\partial_u + \wD_x (f)\partial_{u_1}+\wD_x^2
(f)\partial_{u_2} + \ldots 
\end{equation}
where $f \in C^{\infty}(Y)$  are such that
\begin{equation}\label{sec1:eq:sym}
        \ell_F(f) = 0.
\end{equation}
Here, $\ell_F$ is the \emph{universal linearisation
operator} \cite{AMV-local,KLV} which is Fr\'echet derivative of
$F$, and it reads in the case of the KdV-equation \eqref{sec1:eq:kdv}
\begin{equation}\label{sec1:eq:sym1}
        \wD_t(f) -u \wD_x(f) - u_1\cdot f - (\wD_x)^3(f) = 0.
\end{equation}

Let  now $W\subset \BBR^m$ with coordinates $(w_1, \cdots w_m)$.\\
The Cartan distribution on $Y \otimes W$ is given by
\begin{align}
\begin{array}{lclcl}
\overline{D}_x &=& \wD_x &+& \sum\limits_{j=1}^m X^j\partial_{w_j},\\
\overline{D}_t &=& \wD_t &+& \sum\limits^m_{j=1} T^j\partial_{w_j}
\end{array}
\end{align}
where $X^j,T^j \in C^\infty(Y\otimes W)$ such that
\begin{equation}\label{sec1:eq:com}
        [\oD_x,\oD_t] = 0
\end{equation}
which yields the so called \emph{covering condition}
\[
        \wD_x(T) - \wD_t(X) + [X,T] = 0
\]
whereas in \eqref{sec1:eq:com} [*,*] is the Lie bracket for vector fields
$X=\sum^m_{j=1} X^j\partial_{w_j}, ~T = \sum^m_{j=1}T^j\partial_{w_j}$
defined on $W$.

A \emph{nonlocal symmetry} is a vertical vector field on $Y \otimes W$,
i.e.\ of the form  \eqref{sec1:eq:vec}, which satisfies $(f \in
C^\infty(Y \otimes W))$
\begin{equation}\label{sec1:eq:nonlsym}
        \bar{\ell}_F(f) = 0
\end{equation}
which for the KdV-equation results in
\begin{equation}
        \oD_t(f) - u\oD_x(f) - u_1\cdot f - (\oD_x)^3(f) = 0.
\end{equation}
Formally, this is just what is called the \emph{shadow} of the symmetry,
i.e.,
not bothering about the ${\partial_{w^j}}$ $(j=1 \ldots m)$ components.\\
In effect the full symmetry should also satisfy the invariance of the 
equations governing the nonlocal variables $w_j$ $(j=1,\ldots,m)$; i.e.,
\begin{align}
(w_j)_x&=X^j,\nonumber\\
(w_j)_t&=T^j.\nonumber
\end{align}
The construction of the associated ${\partial_{w^j}}$ $(j=1,
  \ldots, m)$ components is called the \emph{reconstruction problem} 
\cite{KV-book}.
For reasons of simplicity, we omit this reconstruction problem, i.e.
reconstructing the complete vector field or full symmetry from its shadow.\\
The classical Lenard recursion operator $\CR$ for the KdV-equation,
\begin{equation}
        \CR = D^2_x + {2\over 3}u + {1\over 3}u_{1}D^{-1}_x
\end{equation}
which is just such, that
\newpage
\begin{align}
        f_0 &= u_1,\nonumber\\
        \CR f_0 = f_1 &= uu_1 + u_{3},\\
        \CR f_1 = f_2 &= u_{5} + {5 \over 3}u_{3}u + {10\over
                                         3}u_{2}u_{1} + {5 \over
6}u_{1}u^2,\nonumber
\end{align}
i.e., creating the $(x,t)$--independent hierarchy of higher
symmetries, has an action on the vertical symmetry $\re_{\of_{-1}}$
(Gallilei-boost)
\begin{align}
        \of_{-1} &= (1+tu_{1})/3,\nonumber\\
        \CR\of_{-1} &= \of_0 = 2u +  xu_{1} + 3t(u_{3} +
uu_{1}),\\
        \of_1 = \CR\of_0 &= 3t f_2 + x f_1 + 4u_{2} + {4 \over 3}
u^2 + {1\over 3}u_{1} D^{-1}_x(u).\nonumber       
\end{align}
If we introduce the variable $p$ $(=w_1)$ through
\begin{align}
        p_x &= u,\nonumber\\
        p_t &= u_{2} + {1\over 2}u^2,\\
        \mbox{i.e.} \quad D_t(u) &= D_x(u_{2} + {1\over 2}u^2),\nonumber
\end{align}
then $\re_{\of_1}$ is the shadow of a nonlocal symmetry in the
one-dimensional covering of the KdV-equation by 
\[
p = w_1, ~ X_1 = u , ~T_1= u_{2} + {1\over 2}u^2.
\]
So, by its action the Lenard recursion operator creates nonlocal
symmetries in a natural way.\\
More applications of nonlocal symmetries can be found in
e.g.\cite{KrasKers-book}.
\subsection{A Special Type of Covering: the Cartan-covering}
We discuss a special type of the nonlocal setting indicated in the
previous section, the so called Cartan-covering. As mentioned before we
shall illustrate this by the KdV-equation.\\
Let $Y \subset J^\infty(x,t;u)$ be the infinite prolongation of the
KdV-equation \eqref{sec1:eq:kdv}. Contact one forms on $TJ^\infty(x,t;u)$
are given by
\begin{align}\label{sec2:contforms}
\alpha_0 &= du - u_{1}dx - u_tdt,\nonumber\\
\alpha_1 &= du_{1} - u_{2}dx - u_{1t}dt,\\
\alpha_2 &= du_{2} - u_{3}dx - u_{2t}dt,~~ ... ~~.\nonumber
\end{align}
From the total partial derivative operators of the previous section we
have
\[
\wD_x(\alpha_0) = \alpha_1, 
~ \wD_x(\alpha_1) =
\alpha_2,~\ldots ,~ \wD_x(\alpha_i) =
\alpha_{i+1},~\ldots ,
\]
\begin{equation}
\begin{array}{rcl}
\wD_t (\alpha_0)&=& \alpha_0u_x + \alpha_1u + \alpha_3 = \alpha_t ,\\
\wD_t(\alpha_i) &=& (\wD_x)^i(\alpha_t).
\end{array}
\end{equation}
We now define the Cartan-covering of $Y$ by $Y \otimes \BBR^\infty$ 
\begin{align}\label{sec2:cartan}
D^C_x &= \wD_x + \sum_i(\alpha_{i+1}){\partial\over
\partial\alpha_i},\nonumber\\
D^C_t &= \wD_t + \sum_i(\wD_x)^i\alpha_t{\partial\over
\partial\alpha_i}
\end{align}
where local coordinates are given
$(x,t,u,u_{1},\ldots,\alpha_0,\alpha_1,\ldots)$. 

It is a straightforward check, and obvious that
\begin{equation}
        [D^C_x,D^C_t] = 0,
\end{equation}
i.e.\ they form a Cartan distribution  on $Y \otimes \BBR^\infty$.\\
{\bf Note 1:}\\
Since at first $\alpha_i$ ($i = 0,\ldots$) are contact forms, they
constitute a Grassmann algebra (graded commutative algebra)
$\Lambda(\alpha)$, where
\begin{equation}
        \alpha_i \wedge \alpha_j = -\alpha_j \wedge \alpha_i,\nonumber
\end{equation}
i.e.,
\[
        xy = (-1)^{|x||y|} yx
\]
where $x,y$ are contact $(*)$-forms of degree $|x|$ and $|y|$,
respectively.
So  in effect we are dealing with a \emph{graded} covering.\\
{\bf Note 2:}\\
Once we have introduced the Cartan-covering by \eqref{sec2:cartan}  we
can forget
about the specifics of $\alpha_i$ ($i = 0,1,\ldots$) and just treat them
as (odd) ordinary variables, associated with their differentiation
rules.\\
One can discuss nonlocal symmetries in this type of covering just as
in the previous section, the only difference being:
\[
f\in C^{\infty} (Y ) \otimes \wedge (\alpha).
\]
In the next section we shall combine constructions of the previous
subsection and this one, in order to construct the recursion operator
for symmetries.

\subsection{The Recursion Operator as Symmetry in the Cartan-covering}
We shall discuss the recursion operator for symmetries of
the KdV-equation as a geometrical object, i.e., a symmetry in the
Cartan-covering.\\
Our starting point is the four dimensional covering of the KdV-equation in
$Y \otimes\BBR^4$ where
\begin{align}\label{sec3:cov}
        \oD_x & = {\widetilde D}_x + u\partial_{w_1} + {1\over 2}
        u^2\partial_{w_2} + (u^3-3u^2_1)\partial_{w_3} +
        w_1\partial_{w_4},\nonumber\\
        \oD_t & =  {\widetilde D}_t + ({1\over 2}u^2 + u_2)
\partial_{w_1} + ({1\over 3}
        u^3 - {1\over 2} u^2_1 + uu_2)\partial_{w_2}\\
        & +  ({3\over 4}u^4 - 6u_1u_3 + 3u^2u_2 - 6uu_1^2 + 3u^2_2)
\partial_{w_3} +
        (u_1 + w_2)\partial_{w_4}.\nonumber
\end{align}
$\oD_x,\oD_t$ satisfy the covering condition \eqref{sec1:eq:com}, and
note, that due
to the fact that the coefficients of $\partial_{w_i}$ ($i =1,2,3$) in
\eqref{sec3:cov} are independent of $w_j$ ($j =1,2,3$), these coefficients
constitute {\bf local} conservation laws for the KdV-equation. \\
The coefficients $ w_1$ and $ u_1 + w_2$ of $\partial_{w_4}$
constitute the first {\bf nonlocal} conservation law of the 
KdV-equation.\\
We have the following ``formal'' variables:
\newpage
\begin{align}\label{sec3:nonlocaldef}
w_1 &= \int u dx,\nonumber\\
w_2 &= \int {1\over 2} u^2dx,\nonumber\\
w_3 &= \int (u^3-3u_1^2)dx,\\
w_4 &= \int w_1dx \nonumber
\end{align}
where $w_4$ is of a higher or deeper nonlocality.

We now build the Cartan-covering of the previous section on the
covering  given by \eqref{sec3:cov} by introduction of the contact forms
$\alpha_0,\alpha_1,\alpha_2,\ldots$ \eqref{sec2:contforms} and
\begin{align}\label{sec3:nonlcontforms}
        \alpha_{-1} &= dw_1 - udx - ({1\over 2}u^2 + u_2)dt,\nonumber\\
        \alpha_{-2} &= dw_2 - {1\over 2}u^2dx - ({1\over 3}u^3 -
{1\over 2} u^2_1 + uu_2)dt
\end{align}
as well as similarly for $\alpha_{-3}, \alpha_{-4}$.
It is straightforward to prove the following relations
\begin{align}
\oD_x(\alpha_{-1}) = \alpha_0,\ \ \ & \oD_t(\alpha_{-1}) = u\alpha_0 +
\alpha_0,\nonumber\\
\oD_x(\alpha_{-2}) = u\alpha_0,\ \ \ & \oD_t(\alpha_{-2}) = u^2\alpha_0 -
u_1\alpha_1 + u\alpha_2 + u_2\alpha_0,\\
\oD_x(\alpha_{-3}) = 3u^2\alpha_0-6u_1\alpha_1,\ \ \  &\ldots ~.
\nonumber
\end{align}
We are now constructing symmetries in this Cartan-covering of
the KdV-equation which are linear w.r.t.  $\alpha_i$ ($i =
-4,\ldots,0,1,\ldots$).

The symmetry condition for $f \in C^\infty (Y \otimes \BBR^4) \otimes
\Lambda^1(\alpha)$ is just given by \eqref{sec1:eq:sym1}
\begin{equation}\label{sec1:eq:cartsym}
        \bar{\ell}_F^C(f) = 0
\end{equation}
which for the KdV-equation results in
\[
        \oD_t^C(f) - u\oD^C_x(f) - u_xf - (\oD^C_x)^3f = 0.
\]
As solutions of these equations we obtain
\begin{equation}\label{kdv:cartsym}
\begin{align}
        f^0 &= \alpha_0,\nonumber\\
        f^1 &= ({2\over 3}u) \alpha_0 + \alpha_2 + ({1\over 3}u_1)
                \alpha_{-1},\\
        f^2 &= ({4\over 9}u^2 + {4\over 3}u_2)\alpha_0 + (2u_1)
\alpha_1 + ({4\over 3}u )\alpha_2 + \alpha_4\nonumber\\
& + {1\over 3}(uu_1 + u_3)\alpha_{-1} + {1\over 9} (u_1)
\alpha_{-2}.\nonumber
\end{align}
\end{equation}
As we mentioned above we are working  in effect with form-valued
vector fields $\re_{f^0},\re_{f^1}\re_{f^2}$.
For these objects one can define
Fr\"olicher-Nijenhuis and (by contraction) Richardson-Nijenhuis
brackets \cite{kr1},\cite{KrasKers-book}
Without going into details, for which the reader is referred to
\cite{kr1},
we can construct the contraction of a (generalized) symmetry  and a
form-valued symmetry e.g.
\begin{equation}
        R = ({2\over 3}u\alpha_0 + \alpha_2 + {1\over
3}u_1\alpha_{-1}){\partial \over \partial u} + \ldots ~.
\end{equation}
The contraction, formally defined by 
\begin{equation}
V\lrcorner
R_u=\sum_\alpha V_\alpha R_u^\alpha  \nonumber
\end{equation}
where $\alpha$ runs over all local
and nonlocal variables, is given by
\begin{equation}
        (V\lrcorner R) = (V \lrcorner R_u)\partial_u + \oD^C_x(V
\lrcorner R_u) \partial_{u_1} + \ldots ~.
\end{equation}
Start now with
\begin{equation}
        V_1 = u_1 {\partial\over \partial u} + u_2 {\partial \over
\partial u_1} + \ldots
\end{equation}
whose prolongation in the setting $Y \otimes \BBR^4$ is
\begin{align}
\overline{V}_1 = u_1{\partial\over \partial u} + u_2{\partial\over
\partial u_1} + \ldots + u{\partial\over \partial w_1} + {1\over
2}u^2{\partial\over \partial w_2} & + 
 (u^3 - 3u^2_1) {\partial\over
\partial w_3}\nonumber\\ &+ w_1 {\partial\over \partial w_4}
\end{align}
then
\begin{align}
(\overline{V}_1 \lrcorner R) & =  [({2\over 3}u)u_1 + 1\cdot u_3 + {1\over 3}
u_1 \cdot u]{\partial\over \partial u} + \ldots\nonumber\\
&= (u_3+uu_1) {\partial\over \partial u} + \ldots = V_3
\end{align}
and similarly
\begin{equation}
        (\overline{V}_3 \lrcorner R) = (u_5 + {5\over 3}u_3u + {10\over
3} u_2u_1 + {5\over 6}u^2u_1){\partial\over \partial u} + \ldots = V_5.
\end{equation}
The result given above means that the well known Lenard recursion
operator for symmetries of the KdV-equation is represented as a
\emph{symmetry}, $\re_{f_{1}}$ , in the Cartan-covering of this
equation and in effect is a geometrical object.
\newpage

\section{The $N=2$ $ \alpha=1$ Supersymmetric KdV-equation}\label{sec:2}
In this section we shall discuss all computations leading at the end
to the complete integrability of the $N=2$  $\alpha=1$ supersymmetric
KdV-equation.\\
The $N=2$  $\alpha=1$ supersymmetric KdV-equation is described by
\begin{equation}\label{sec2:eq:kdv}
J_t = \{ J_{zz}+3J[D,\oD]J+J^3\}_z
\end{equation}
in the $N=2$ superspace with a coordinate $Z=(z,\theta,\oth)$,\ $_z$
denotes derivative with respect to $z$ and $ D, \oD$ are the fermionic 
covariant  derivatives of the $N=2$ supersymmetry, governed by definitions
\begin{eqnarray}
D=\frac{\partial}{\partial\theta}
 -\frac{1}{2}\overline\theta\frac{\partial}{\partial z}, \quad
{\overline D}=\frac{\partial}{\partial\overline\theta}
 -\frac{1}{2}\theta\frac{\partial}{\partial z}, \quad
D^{2}={\overline D}^{2}=0, \quad
\left\{ D,{\overline D} \right\}= -\frac{\partial}{\partial z}
\equiv -{\partial}.\nonumber
\label{DD}
\end{eqnarray}
Formally, the nonlocal setting for differential equations of the previous
section was done for classical equations, but applies too for
supersymmetric equations \cite{KrasKers-book}.\\
In order to discuss conservation laws, symmetries  and deformations we 
choose local coordinates in the infinite jet bundle
$Y((z,\theta,\oth);J)$, where we choose as local even coordinates
\begin{equation}\label{sec2:eq:coev}
z,t,J,D \oD J,J_{z},D \oD J_{z},J_{zz},D \oD J_{zz},J_{zzz},\ldots\nonumber
\end{equation}
and odd coordinates
\begin{equation}\label{sec2:eq:coodd}
\theta,\oth,D J,\oD J,D J_{z},\oD J_{z},D J_{zz},\oD J_{zz},\ldots 
~.\nonumber
\end{equation}
In order to  have a complete setting we first describe the
construction of conservation laws and the associated introduction of
nonlocal variables.\\
In effect we construct an abelian covering of the equation
structure.\\
In the first subsection we shall discuss conservation laws and the
associated nonlocalities.\\
In the next subsection we obtain nonlocal symmetries associated to these
nonlocalities, which turn out to be $(\theta,\oth)$ - dependent and arise
in so called quadruplets, similar to the
nonlocalities.\\
In the subsection 3.3  we derive an explicit deformation of
the equation structure leading to the recursion operator for symmetries.\\
Finally in the last subsection  we obtain the factorization of the
recursion operator as product of the second Hamiltonian operator  and
the inverse of the zero-th Hamiltonian operator.

\subsection{Conservation laws and nonlocal variables}\label{sec:2.1}
Here we shall construct conservation laws for (\ref{sec2:eq:kdv})
in order to arrive at an abelian covering.\\ 
So we construct $X=X(z,t,J,\ldots), ~T=T(z,t,J,\ldots)$
such that
\begin{align}
D_z(T)=D_t(X)
\end{align}
and in a similar way we construct nonlocal conservation laws by the
requirement
\begin{align}
\bar{D}_z(\bar{T})=\bar{D}_t(\bar{X})
\end{align}
where $\bar{D}_z,\bar{D}_t$ are defined by
\begin{align}
\bar{D}_z=\partial_z&+J_{z}\partial_{J}+D \oD J_{z}\partial_{D \oD
J}+\ldots\nonumber\\
&+D J_{z}\partial_{D J}+\oD J_{z}\partial_{\oD J}+\ldots\nonumber\\
\bar{D}_t=\partial_t&+J_{t}\partial_{J}+D \oD J_{t}\partial_{D \oD
J}+\ldots\nonumber\\
&+D J_{t}\partial_{D J}+\oD J_{t}\partial_{\oD J}+\ldots ~.
\end{align}
Moreover $\bar{X}$,$\bar{T}$ are dependent on local variables $z$,
$t$, $J,\dots,$ as well as the already determined nonlocal variables,
denoted here by $p_*$, which are associated to the conservation laws
$(X,T)$ by the
formal definition
\begin{align*}
D_z(p_{*})&=(p_{*})_z=X,\\
D_t(p_{*})&=(p_{*})_t=T.
\end{align*}
Proceeding in this way, we obtained a number of conservation laws,
which arise as multiplets of four conservation laws each. The 
corresponding nonlocal variables are
\begin{align}
\label{kdv0}
P_0, ~~&D P_0, ~~~\oD P_0,\quad  ~~~~D \oD P_0,\nonumber \\
Q_{1\over 2}, ~~&D Q_{1\over 2}, ~~\oD Q_{1\over 2}, ~~D \oD Q_{1\over
2},\nonumber \\
\oQ_{1\over 2}, ~~&D \oQ_{1\over 2}, ~~\oD \oQ_{1\over 2},~~D \oD
\oQ_{1\over 2},\\
P_1, ~~&D P_1, ~~~~~~\oD P_1,\quad ~~~~ D \oD P_1 \nonumber
\end{align}
where their defining equations are given by
\begin{align}\label{kdv1}
(P_0)_z &=J,\nonumber\\
(Q_{1\over 2})_z &=e^{(+)} D J,\nonumber\\
(\oQ_{1\over 2})_z &=e^{(-)} \oD J,\\
(P_1)_z &=e^{(+)}(\oD J)(D P_0)+e^{(-)}(\oD J)(Q_{1\over 2}).\nonumber
\end{align}\\
In (\ref{kdv1}) $e^{(+)}$ and $e^{(-)}$ refer to $e^{(+2P_0)}$ and
$e^{(-2P_0)}$ respectively.\\
It should be noted that the quadru-plet $P_0$ satisfies differentiation
rules as follows:
\begin{align}\label{kdv2}
D(P_0)&=D P_0,\nonumber\\
D(D P_0)&=0,\nonumber\\
D(\oD P_0)&=D \oD P_0,\nonumber\\
D(D \oD P_0)&=0,\nonumber\\
  &  \\
\oD(P_0)&=\oD P_0,\nonumber\\
\oD(D P_0)&=- (P_0)_{z}-D \oD P_0,\nonumber\\
\oD(\oD P_0)&=0,\nonumber\\
\oD(D \oD P_0)&=-\oD(P_0)_z,\nonumber
\end{align}\\
and similarly for other quadru-plets $Q_{1\over 2}, \oQ_{1\over 2},
P_1$.\\
It should be noted that for the two other $N=2$ KdV-hierarchies
($\alpha$=4 and \\-2), despite their original $N=2$ supersymmetric
structure, their conservation laws do not form supersymmetric multiplets.\\
So in effect we have at this moment sixteen local and nonlocal
conservation laws leading to a similar number of new nonlocal
variables.\\
They perfectly match with those ones obtained previously
\cite{KrasKers-book} (page 340, eq. (7.78)).\\
If we arrange them according to their respectively degrees we arrive
at:
\newpage
\begin{align*}
0: ~ & P_0;\\
{1\over 2}: ~ & D P_0, \oD P_0, Q_{1\over 2}, \oQ_{1\over 2}; \\
1: ~ & D \oD P_0,D Q_{1\over 2}, D \oQ_{1\over 2} ,\oD Q_{1\over 2}, \oD
\oQ_{1\over 2} , P_1;\\
{3\over 2}: ~ &D \oD Q_{1\over 2}, D \oD \oQ_{1\over 2}, D P_1,\oD P_1;\\
2: ~ & D \oD P_1.
\end{align*}\\
In Subsection \ref{sec:2.2} we shall discuss  local and nonlocal
symmetries of eq. (\ref{sec2:eq:kdv})
while in Subsection \ref{sec:2.3} we construct the recursion operator or
deformation of the equation structure.

\subsection{Local and nonlocal symmetries}\label{sec:2.2}
In this section we shall present results for the construction of local and
nonlocal symmetries of (\ref{sec2:eq:kdv}). In order to construct
these symmetries, we consider the system of partial differential equations
obtained by the infinite prolongation.\\
First we present the {\bf local} symmetries as they are required for explicit
formulae for the coefficients which arise in the form-valued vector
field of the  next subsection leading to the recursion operator  for
symmetries.\\

$Y_{1}:=J_{z},$\\

$Y_{3}:= \{ J_{zz}+3J[D,\oD]J+J^3\}_z, $\\

$Y_{5} :=   - 10\cdot \oD J\cdot D J_{zz}\cdot J - 20\cdot \oD J\cdot D
J_{z}\cdot J^2 - 10\cdot \oD J\cdot D J_{z}\cdot J_{z} - 10\cdot 
D J\cdot \oD J_{zz}\cdot J + 20\cdot D J\cdot \oD J_{z}\cdot J^2 -
10\cdot D J\cdot \oD J_{z}\cdot J_{z} + 40\cdot D J\cdot \oD J\cdot 
J\cdot J_{z} + 5\cdot J^4\cdot J_{z} + 20\cdot J^3\cdot D \oD J_{z} +
10\cdot J^3\cdot J_{zz} + 60\cdot J^2
\cdot D \oD J\cdot J_{z} + 30\cdot J^2\cdot J_{z}^2 + 10\cdot J^2\cdot
J_{zzz} + 80\cdot J\cdot D \oD J\cdot D \oD J_{z} 
+ 40\cdot J\cdot D \oD J\cdot J_{zz} + 5\cdot J\cdot J_{zzzz} + 40\cdot
J\cdot J_{z}\cdot D \oD J_{z} + 50\cdot J\cdot J_{z}\cdot J_{zz} 
+ 10\cdot J\cdot D \oD J_{zzz} + 40\cdot D \oD J^2\cdot J_{z} + 40\cdot D
\oD J\cdot J_{z}^2 + 10\cdot D \oD J\cdot 
J_{zzz} + J_{zzzzz} + 15\cdot J_{z}^3 + 20\cdot J_{z}\cdot D \oD J_{zz} +
15\cdot J_{z}\cdot J_{zzz} + 20\cdot 
D \oD J_{z}\cdot J_{zz} + 10\cdot J_{zz}^2,$\\

$Y_{7} := ( - 42\cdot \oD J_{z}\cdot D J_{zzz}\cdot J - 126\cdot \oD
J_{z}\cdot D J_{zz}\cdot J^2 - 28\cdot \oD J_{z}\cdot D J_{zz}\cdot 
J_{z} - 
42\cdot D J_{z}\cdot \oD J_{zzz}\cdot J + 126\cdot D J_{z}\cdot \oD
J_{zz}\cdot J^2 - 28\cdot D J_{z}\cdot \oD J_{zz}\cdot J_{z} + 364\cdot D
J_{z}\cdot \oD J_{z}\cdot J\cdot J_{z} 
- 28\cdot \oD J\cdot D J_{zzzz}\cdot J - 70\cdot \oD J\cdot D
J_{zzz}\cdot J^2 - 42\cdot \oD J\cdot D J_{zzz}\cdot J_{z} - 70\cdot \oD
J\cdot D J_{zz}\cdot J^3 -
168\cdot \oD J\cdot D J_{zz}\cdot J\cdot D \oD J - 350\cdot \oD J\cdot D
J_{zz}\cdot J\cdot J_{z} - 28\cdot \oD J\cdot D J_{zz}\cdot J_{zz} -
84\cdot \oD J\cdot D J_{z}\cdot 
J^4 - 336\cdot \oD J\cdot D J_{z}\cdot J^2\cdot D \oD J - 378\cdot \oD
J\cdot D J_{z}\cdot J^2\cdot J_{z} - 168\cdot \oD J\cdot D J_{z}\cdot
J\cdot 
D \oD J_{z} - 294\cdot \oD J\cdot D J_{z}\cdot J\cdot J_{zz} - 168\cdot
\oD J\cdot D J_{z}\cdot D \oD J\cdot J_{z} - 238\cdot \oD J\cdot D
J_{z}\cdot J_{z}^2 - 
14\cdot \oD J\cdot D J_{z}\cdot J_{zzz} - 28\cdot D J,12)\cdot J +
70\cdot D J\cdot \oD J_{zzz}\cdot J^2 - 42\cdot D J\cdot \oD J_{zzz}\cdot
J_{z} -
70\cdot D J\cdot \oD J_{zz}\cdot J^3 - 168\cdot D J\cdot \oD J_{zz}\cdot
J\cdot D \oD J + 182\cdot D J\cdot \oD J_{zz}\cdot J\cdot J_{z} - 28\cdot
D J\cdot \oD J_{zz}
\cdot J_{zz} + 84\cdot D J\cdot \oD J_{z}\cdot J^4 + 336\cdot D J\cdot
\oD J_{z}\cdot J^2\cdot D \oD J - 42\cdot D J\cdot \oD J_{z}\cdot
J^2\cdot J_{z} -
168\cdot D J\cdot \oD J_{z}\cdot J\cdot D \oD J_{z} + 126\cdot D J\cdot
\oD J_{z}\cdot J\cdot J_{zz} - 168\cdot D J\cdot \oD J_{z}\cdot D \oD
J\cdot J_{z} + 70\cdot 
D J\cdot \oD J_{z}\cdot J_{z}^2 - 14\cdot D J\cdot \oD J_{z}\cdot J_{zzz}
+ 336\cdot D J\cdot \oD J\cdot J^3\cdot J_{z} + 336\cdot D J\cdot \oD
J\cdot 
J^2\cdot D \oD J_{z} + 168\cdot D J\cdot \oD J\cdot J^2\cdot J_{zz} +
672\cdot D J\cdot \oD J\cdot J\cdot D \oD J\cdot J_{z} + 336\cdot D
J\cdot \oD J\cdot J\cdot J_{z}^2 + 84\cdot D J\cdot \oD J\cdot J\cdot
J_{zzz} + 140\cdot D J\cdot \oD J\cdot J_{z}\cdot 
J_{zz} + 7\cdot J^6\cdot J_{z} 
+ 42\cdot J^5\cdot D \oD J_{z} + 21\cdot J^5\cdot J_{zz} + 210\cdot
J^4\cdot D \oD J\cdot J_{z} + 105\cdot J^4\cdot J_{z}^2 + 35
\cdot J^4\cdot J_{zzz} + 504\cdot J^3\cdot D \oD J\cdot D \oD J_{z} +
252\cdot J^3\cdot D \oD J\cdot J_{zz} + 35\cdot J^3\cdot J_{zzzz} + 
252\cdot J^3\cdot J_{z}\cdot D \oD J_{z} + 350\cdot J^3\cdot J_{z}\cdot
J_{zz} + 70\cdot J^3\cdot D \oD J_{zzz} + 756\cdot J^2\cdot D \oD J
^2\cdot J_{z} + 756\cdot J^2\cdot D \oD J\cdot J_{z}^2 + 224\cdot
J^2\cdot D \oD J\cdot J_{zzz} + 21\cdot J^2\cdot J_{zzzzz} + 315\cdot 
J^2\cdot J_{z}^3 + 406\cdot J^2\cdot J_{z}\cdot D \oD J_{zz} + 315\cdot
J^2\cdot J_{z}\cdot J_{zzz} + 420\cdot J^2\cdot D \oD J_{z}\cdot 
J_{zz} + 210\cdot J^2\cdot J_{zz}^2 + 840\cdot J\cdot D \oD J^2\cdot D
\oD J_{z} + 420\cdot J\cdot D \oD J^2\cdot J_{zz} + 98\cdot J
\cdot D \oD J\cdot J_{zzzz} + 840\cdot J\cdot D \oD J\cdot J_{z}\cdot D
\oD J_{z} + 1148\cdot J\cdot D \oD J\cdot J_{z}\cdot J_{zz} + 196\cdot
J\cdot D \oD J\cdot 
D \oD J_{zzz} + 147\cdot J\cdot J_{zzzz}\cdot J_{z} + 14\cdot J\cdot D
\oD J_{zzzzz} + 7\cdot J\cdot J_{zzzzzz} + 742\cdot J\cdot J_{z}^2\cdot D
\oD J_{z} + 735
\cdot J\cdot J_{z}^2\cdot J_{zz} + 98\cdot J\cdot J_{z}\cdot D \oD
J_{zzz} + 420\cdot J\cdot D \oD J_{z}\cdot D \oD J_{zz} + 210\cdot J\cdot
D \oD J_{z}\cdot 
J_{zzz} + 210\cdot J\cdot J_{zz}\cdot D \oD J_{zz} + 245\cdot J\cdot
J_{zz}\cdot J_{zzz} + 280\cdot D \oD J^3\cdot J_{z} + 420\cdot D \oD J^
2\cdot J_{z}^2 + 84\cdot D \oD J^2\cdot J_{zzz} + 14\cdot D \oD J\cdot
J_{zzzzz} + 350\cdot D \oD J\cdot J_{z}^3 + 364\cdot D \oD J\cdot
J_{z}\cdot 
D \oD J_{zz} + 266\cdot D \oD J\cdot J_{z}\cdot J_{zzz} + 336\cdot D \oD
J\cdot D \oD J_{z}\cdot J_{zz} +
168\cdot D \oD J\cdot J_{zz}^2 + 42\cdot  J_{zzzz}
\cdot D \oD J_{z} + 56\cdot J_{zzzz}\cdot J_{zz} + 42\cdot D \oD
J_{zzzz}\cdot J_{z} + 28\cdot J_{zzzzz}\cdot J_{z} + J_{zzzzzzz} +
105\cdot J_{z}^4 + 182\cdot  J_{z}
^2\cdot D \oD J_{zz} + 210\cdot J_{z}^2\cdot J_{zzz} + 280\cdot
J_{z}\cdot D \oD J_{z}^2 +
448\cdot J_{z}\cdot D \oD J_{z} \cdot J_{zz} + 280\cdot 
J_{z}\cdot J_{zz}^2 + 70\cdot J_{zz}\cdot D \oD J_{zzz} + 70\cdot D \oD
J_{zz}\cdot J_{zzz} + 35\cdot J_{zzz}^2).$\\

It should be noted that the symmetries $Y_{5}$ and $Y_{7}$ are rather 
massive, containing 28 and 104 terms respectively.
We made our choice for a presentation as shown above, in order to have 
expressions, which are quite huge, as readable as possible.\\

Now, we present the four multiplets of nonlocal 
$(\theta,{\bar \theta})$--dependent symmetries as they were constructed
in the above described nonlocal setting.\\

The nonlocal $(\theta, \oth)$--dependent symmetries at level
$0,1/2,1,3/2,2$ are given as follows:\\
This represents the first quadru-plet associated to $P_0$\\
$Y_{P_0,\theta \oth} := \theta \oth\cdot J_{z} + \theta\cdot D J -
\oth\cdot \oD J,$\\

$Y_{P_0, \oth} := \oth\cdot J_{z} + D J,$\\

$Y_{P_0,\theta} := \theta\cdot J_{z} + \oD J,$\\

$Y_{P_0} := J_{z}$.\\

The second quadru-plet associated to $Q_{1\over 2}$ is represented by\\
$Y_{Q_{1\over 2},\theta \oth} := (2\cdot \theta \oth\cdot e^{(+)}\cdot (
- 2\cdot D P_0\cdot D \oD J - 2\cdot D P_0\cdot J_{z} - D J_{z} + 2\cdot
D J\cdot D P_0\cdot \oD P_0
+ D J\cdot D \oD P_0) + 2\cdot \theta \oth\cdot (Q_{1\over 2}\cdot J_{z}
+ \oD J\cdot D Q_{1\over 2} + D J\cdot \oD Q_{1\over 2}
) - 2\cdot \theta\cdot e^{(+)}\cdot D J\cdot D P_0 + 2\cdot \theta\cdot D
J\cdot Q_{1\over 2} - 2\cdot \oth\cdot e^{(+)}\cdot (D J\cdot \oD P_0 + D
\oD J + 
J_{z}) - 2\cdot \oth\cdot \oD J\cdot Q_{1\over 2} - e^{(+)}\cdot D
J)/2,$\\

$Y_{Q_{1\over 2},\oth} := \oth\cdot e^{(+)}\cdot ( - 2\cdot D P_0\cdot D
\oD J - 2\cdot D P_0\cdot J_{z} - D J_{z} + 2\cdot D J\cdot D P_0\cdot
\oD P_0 + 
D J\cdot D \oD P_0) + \oth\cdot (Q_{1\over 2}\cdot J_{z} + \oD J\cdot D
Q_{1\over 2} + D J\cdot \oD Q_{1\over 2}) - e^{(+)}
\cdot D J\cdot D P_0 + D J\cdot Q_{1\over 2},$\\

$Y_{Q_{1\over 2},\theta} := \theta\cdot e^{(+)}\cdot ( - 2\cdot D
P_0\cdot D \oD J - 2\cdot D P_0\cdot J_{z} - D J_{z} + 2\cdot D J\cdot D
P_0\cdot \oD P_0 + 
D J\cdot D \oD P_0) + \theta\cdot (Q_{1\over 2}\cdot J_{z} + \oD J\cdot D
Q_{1\over 2} + D J\cdot \oD Q_{1\over 2}) + e^{(+)}\cdot (D J\cdot \oD
P_0 + D \oD J + J_{z}) + \oD J\cdot Q_{1\over 2},$\\

$Y_{Q_{1\over 2}}  := e^{(+)}\cdot ( - 2\cdot D P_0\cdot D \oD J - 2\cdot
D P_0\cdot J_{z} - D J_{z} + 2\cdot D J\cdot D P_0\cdot \oD P_0 + 
D J\cdot D \oD P_0) + Q_{1\over 2}\cdot J_{z} + \oD J\cdot D Q_{1\over 2}
+ D J\cdot \oD Q_{1\over 2}$.\\

The third quadru-plet associated to $\oQ_{1\over 2}$ is represented by\\

$Y_{\oQ_{1\over 2},\theta \oth} := (2\cdot \theta \oth\cdot e^{(+)}\cdot
(\oQ_{1\over 2}\cdot J_{z} + \oD J\cdot D \oQ_{1\over 2} + D J\cdot \oD
\oQ_{1\over 2}) + 2\cdot \theta \oth\cdot (2\cdot 
\oD P_0\cdot D \oD J + \oD J_{z} + 2\cdot \oD J\cdot D P_0\cdot \oD P_0 -
\oD J\cdot J - \oD J\cdot D \oD P_0)
+ 2\cdot \theta\cdot e^{(+)}\cdot D J\cdot \oQ_{1\over 2} + 2\cdot
\theta\cdot (\oD J\cdot D P_0 + D \oD J) - 2\cdot \oth\cdot e^{(+)}\cdot
\oD J\cdot \oQ_{1\over 2} + 2\cdot 
\oth\cdot \oD J\cdot \oD P_0 - \oD J)/(2\cdot e^{(+)}),$\\

$Y_{\oQ_{1\over 2},\oth} := (\oth\cdot e^{(+)}\cdot (\oQ_{1\over 2}\cdot
J_{z} + \oD J\cdot D \oQ_{1\over 2} + D J\cdot \oD \oQ_{1\over 2}) +
\oth\cdot (2\cdot \oD P_0\cdot 
D \oD J
+ \oD J_{z} + 2\cdot \oD J\cdot D P_0\cdot \oD P_0 - \oD J\cdot J - \oD
J\cdot D \oD P_0) + e^{(+)}\cdot D J\cdot \oQ_{1\over 2} + \oD J\cdot D
P_0 + D \oD J)/e^{(+)},$\\

$Y_{\oQ_{1\over 2},\theta }  := (\theta\cdot e^{(+)}\cdot (\oQ_{1\over
2}\cdot J_{z} + \oD J\cdot D \oQ_{1\over 2} + D J\cdot \oD \oQ_{1\over
2}) + \theta\cdot (2\cdot \oD P_0\cdot 
D \oD J +
\oD J_{z} + 2\cdot \oD J\cdot D P_0\cdot \oD P_0 - \oD J\cdot J - \oD
J\cdot D \oD P_0) + e^{(+)}\cdot \oD J\cdot \oQ_{1\over 2}
 - \oD J\cdot \oD P_0)/e^{(+)},$\\

$Y_{\oQ_{1\over 2}}  := (e^{(+)}\cdot (\oQ_{1\over 2}\cdot J_{z} + \oD
J\cdot D \oQ_{1\over 2} + D J\cdot \oD \oQ_{1\over 2}) + 2\cdot \oD
P_0\cdot D \oD J + 
\oD J_{z} + 2\cdot \oD J\cdot D P_0\cdot \oD P_0 - \oD J\cdot J - \oD
J\cdot D \oD P_0)/e^{(+)}.$\\

The fourth quadru-plet associated to $P_1$ is represented by\\

$Y_{P_1,\theta \oth}
:= (\theta \oth\cdot e^{2(+)}\cdot (4\cdot D P_0\cdot \oQ_{1\over 2}\cdot
D \oD J + 4\cdot D P_0\cdot \oQ_{1\over 2}\cdot J_{z} + 2\cdot D
J_{z}\cdot \oQ_{1\over 2} - 
D J\cdot D \oD \oQ_{1\over 2} - 2\cdot D J\cdot \oQ_{1\over 2}\cdot D \oD
P_0 + 2\cdot D J\cdot \oD P_0\cdot D \oQ_{1\over 2} - 4\cdot D J\cdot D
P_0\cdot \oD P_0\cdot \oQ_{1\over 2}
- 2\cdot D J\cdot D P_0\cdot \oD \oQ_{1\over 2} + 2\cdot D \oD J\cdot D
\oQ_{1\over 2} + 2\cdot J_{z}\cdot D \oQ_{1\over 2}) + \theta \oth\cdot
e^{(+)}\cdot ( - 2\cdot 
Q_{1\over 2}\cdot \oQ_{1\over 2}\cdot J_{z} - 2\cdot \oD J\cdot
\oQ_{1\over 2}\cdot D Q_{1\over 2} + 2\cdot \oD J\cdot Q_{1\over 2}\cdot
D \oQ_{1\over 2} - 2\cdot D J\cdot \oQ_{1\over 2}\cdot \oD Q_{1\over 2} +
2\cdot D J\cdot Q_{1\over 2}\cdot \oD \oQ_{1
\over 2} - D J\cdot \oD J + 2\cdot J\cdot J_{z} + 2\cdot D \oD J_{z} +
J_{zz})
 + \theta \oth \cdot (4\cdot \oD P_0\cdot Q_{1\over 2}\cdot D \oD J +
2\cdot \oD J_{z}\cdot Q_{1\over 2} + \oD J\cdot D \oD Q_{1\over 2} -
2\cdot \oD J\cdot Q_{1\over 2}\cdot J 
- 2\cdot \oD J\cdot Q_{1\over 2}\cdot D \oD P_0 + 2\cdot \oD J\cdot \oD
P_0\cdot D Q_{1\over 2} + 4\cdot \oD J\cdot D P_0\cdot \oD P_0\cdot
Q_{1\over 2} - 2\cdot \oD J\cdot D P_0\cdot \oD Q_{1\over 2} - 2\cdot D
\oD J\cdot \oD Q_{1\over 2}) + \theta\cdot 
e^{2(+)}\cdot (2\cdot D J\cdot D P_0\cdot \oQ_{1\over 2} + D J\cdot D
\oQ_{1\over 2}
) + \theta\cdot e^{(+)}\cdot (D P_0\cdot J_{z} + D J_{z} - 2\cdot D
J\cdot Q_{1\over 2}\cdot \oQ_{1\over 2} + D J\cdot J - D J\cdot D \oD
P_0) + \theta\cdot (2\cdot Q_{1\over 2}\cdot D \oD J + 2\cdot \oD J\cdot
D P_0\cdot Q_{1\over 2} - \oD J\cdot D Q_{1\
over 2}) + \oth\cdot 
e^{2(+)}\cdot (2\cdot \oQ_{1\over 2}\cdot D \oD J + 2\cdot \oQ_{1\over
2}\cdot J_{z} + 2\cdot D J\cdot \oD P_0\cdot \oQ_{1\over 2} + D J\cdot
\oD \oQ_{1\over 2})
 + \oth\cdot e^{(+)}\cdot ( - \oD P_0\cdot J_{z} + \oD J_{z} + 2\cdot \oD
J\cdot Q_{1\over 2}\cdot \oQ_{1\over 2} - 2\cdot \oD J\cdot J - \oD
J\cdot D \oD P_0) + \oth\cdot (2\cdot \oD J\cdot \oD P_0\cdot Q_{1\over
2} - \oD J\cdot \oD Q_{1\over 2}) + e^{2(
+)}\cdot D J\cdot \oQ_{1\over 2} + 
e^{(+)}\cdot (\oD J\cdot D P_0 - D J\cdot \oD P_0) - \oD J\cdot Q_{1\over
2})/e^{(+)},$\\

$Y_{P_1,\oth} := (\oth\cdot e^{2(+)}\cdot (4\cdot D P_0\cdot \oQ_{1\over
2}\cdot D \oD J + 4\cdot D P_0\cdot \oQ_{1\over 2}\cdot J_{z} + 2\cdot D
J_{z}\cdot \oQ_{1\over 2} - 
D J\cdot D \oD \oQ_{1\over 2} - 2\cdot D J\cdot \oQ_{1\over 2}\cdot D \oD
P_0 + 2\cdot D J\cdot \oD P_0\cdot D \oQ_{1\over 2} - 4\cdot D J\cdot D
P_0\cdot \oD P_0\cdot \oQ_{1\over 2} - 2\cdot D J\cdot D P_0\cdot \oD
\oQ_{1\over 2} + 2\cdot D \oD J\cdot D 
\oQ_{1\over 2} + 2\cdot J_{z}\cdot D \oQ_{1\over 2}) + \oth\cdot
e^{(+)}\cdot 
( - 2\cdot Q_{1\over 2}\cdot \oQ_{1\over 2}\cdot J_{z} - 2\cdot \oD
J\cdot \oQ_{1\over 2}\cdot D Q_{1\over 2} + 2\cdot \oD J\cdot Q_{1\over
2}\cdot D \oQ_{1\over 2} - 2\cdot D J\cdot \oQ_{1\over 2}\cdot \oD
Q_{1\over 2} + 2\cdot D J\cdot Q_{1\over 2}\cdot
 \oD \oQ_{1\over 2} - D J\cdot \oD J + 2\cdot J\cdot J_{z} + 2\cdot D \oD
J_{z} + 
J_{zz}) + \oth\cdot (4\cdot \oD P_0\cdot Q_{1\over 2}\cdot D \oD J +
2\cdot \oD J_{z}\cdot Q_{1\over 2} + \oD J\cdot D \oD Q_{1\over 2} -
2\cdot \oD J\cdot Q_{1\over 2}\cdot J - 2\cdot \oD J\cdot Q_{1\over
2}\cdot D \oD P_0 + 2\cdot \oD J\cdot \oD P_0\cdot 
D Q_{1\over 2} + 4\cdot \oD J\cdot D P_0\cdot \oD P_0\cdot Q_{1\over 2}
- 2\cdot \oD J\cdot D P_0\cdot \oD Q_{1\over 2} - 2\cdot D \oD J\cdot \oD
Q_{1\over 2}) + e^{2(+)}\cdot (2\cdot D J\cdot D P_0\cdot \oQ_{1\over 2}
 + D J\cdot D \oQ_{1\over 2}) + e^{(+)}\cdot (D P_0\cdot J_{z} + D J_{z}
- 2\cdot D J\cdot Q_{1\over 2}\cdot \oQ_{1\over 2} + 
D J\cdot J - D J\cdot D \oD P_0) + 2\cdot Q_{1\over 2}\cdot D \oD J +
2\cdot \oD J\cdot D P_0\cdot Q_{1\over 2} - 
\oD J\cdot D Q_{1\over 2})/e^{(+)},$\\

$Y_{P_1,\theta} := (\theta\cdot e^{2(+)}\cdot (4\cdot D P_0\cdot
\oQ_{1\over 2}\cdot D \oD J + 4\cdot D P_0\cdot \oQ_{1\over 2}\cdot J_{z}
+ 2\cdot D J_{z}\cdot \oQ_{1\over 2} - 
D J\cdot D \oD \oQ_{1\over 2} - 2\cdot D J\cdot \oQ_{1\over 2}\cdot D \oD
P_0 + 2\cdot D J\cdot \oD P_0\cdot D \oQ_{1\over 2} - 4\cdot D J\cdot D
P_0\cdot \oD P_0\cdot \oQ_{1\over 2} - 2\cdot D J\cdot D P_0\cdot \oD
\oQ_{1\over 2} + 2\cdot D \oD J\cdot D 
\oQ_{1\over 2} + 2\cdot J_{z}\cdot D \oQ_{1\over 2}) + \theta\cdot
e^{(+)}\cdot (
- 2\cdot Q_{1\over 2}\cdot \oQ_{1\over 2}\cdot J_{z} - 2\cdot \oD J\cdot
\oQ_{1\over 2}\cdot D Q_{1\over 2} + 2\cdot \oD J\cdot Q_{1\over 2}\cdot
D \oQ_{1\over 2} - 2\cdot 
D J\cdot \oQ_{1\over 2}\cdot \oD Q_{1\over 2} + 2\cdot D J\cdot Q_{1\over
2}\cdot \oD \oQ_{1\over 2} - D J\cdot \oD J + 2\cdot J\cdot J_{z} +
2\cdot 
D \oD J_{z} + J_{zz}) + \theta\cdot (4\cdot \oD P_0\cdot Q_{1\over
2}\cdot D \oD J + 2\cdot \oD J_{z}\cdot Q_{1\over 2} + \oD J\cdot D \oD
Q_{1\over 2} -
2\cdot \oD J\cdot Q_{1\over 2}\cdot J - 2\cdot \oD J\cdot Q_{1\over
2}\cdot D \oD P_0 + 2\cdot \oD J\cdot \oD P_0\cdot D Q_{1\over 2} +
4\cdot \oD J\cdot D P_0\cdot \oD P_0\cdot Q_{1\over 2} - 2\cdot \oD
J\cdot D P_0\cdot \oD Q_{1\over 2} - 2\cdot D \oD 
J\cdot \oD Q_{1\over 2}) + e^{2(+)}\cdot ( - 2\cdot \oQ_{1\over 2}\cdot D
\oD J - 2\cdot \oQ_{1\over 2}\cdot J_{z} - 2\cdot D J\cdot \oD P_0\cdot
\oQ_{1\over 2} - D J\cdot \oD \oQ_{1\over 2}) + e^{(+)}\cdot (
\oD P_0\cdot J_{z} - \oD J_{z} - 2\cdot \oD J\cdot Q_{1\over 2}\cdot
\oQ_{1\over 2} + 2\cdot \oD J\cdot J + \oD J\cdot 
D \oD P_0) - 2\cdot \oD J\cdot \oD P_0\cdot Q_{1\over 2} + \oD J\cdot \oD
Q_{1\over 2})/e^{(+)},$\\

$Y_{P_1} := (e^{2(+)}\cdot (4\cdot D P_0\cdot \oQ_{1\over 2}\cdot D \oD J
+ 4\cdot D P_0\cdot \oQ_{1\over 2}\cdot J_{z} + 2\cdot D J_{z}\cdot
\oQ_{1\over 2} - D J\cdot D \oD \oQ_{1\over 2} 
- 2\cdot D J\cdot \oQ_{1\over 2}\cdot D \oD P_0 + 2\cdot D J\cdot \oD
P_0\cdot D \oQ_{1\over 2} - 4\cdot D J\cdot D P_0\cdot \oD P_0\cdot
\oQ_{1\over 2} - 2
\cdot D J\cdot D P_0\cdot \oD \oQ_{1\over 2} + 2\cdot D \oD J\cdot D
\oQ_{1\over 2} + 2\cdot J_{z}\cdot D \oQ_{1\over 2}) + e^{(+)}\cdot ( -
2\cdot Q_{1\over 2}\cdot \oQ_{1\over 2}\cdot J_{z} - 2\cdot \oD J\cdot
\oQ_{1\over 2}\cdot D Q_{1\over 2} + 2\cdot
\oD J\cdot Q_{1\over 2}\cdot D \oQ_{1\over 2} - 2\cdot D J\cdot
\oQ_{1\over 2}\cdot 
\oD Q_{1\over 2} + 2\cdot D J\cdot Q_{1\over 2}\cdot \oD \oQ_{1\over 2} -
D J\cdot \oD J + 2\cdot J\cdot J_{z} + 2\cdot D \oD J_{z} + J_{zz}) +
4\cdot \oD P_0\cdot Q_{1\over 2}\cdot D \oD J + 2\cdot \oD J_{z}\cdot
Q_{1\over 2} + \oD J\cdot D \oD Q_{1\over 2} - 2\cdot \oD J\cdot
Q_{1\over 2}\cdot J - 2\cdot \oD J\cdot Q_{1\over 2}\cdot D \oD P_0 +
2\cdot \oD J\cdot \oD P_0\cdot D Q_{1\over 2} + 4
\cdot \oD J\cdot D P_0\cdot \oD P_0\cdot Q_{1\over 2} - 2\cdot \oD J\cdot
D P_0\cdot 
\oD Q_{1\over 2} - 2\cdot D \oD J\cdot \oD Q_{1\over 2})/e^{(+)}.$\\

Note that in previous formulas $e^{2(+)}$ refers to $e^{4P_0}$.

\subsection{Recursion operator}\label{sec:2.3}
Here we present the recursion operator $\CR$ for symmetries for this case
obtained as a higher symmetry in the Cartan covering of equation
(\ref{sec2:eq:kdv}) augmented by equations governing the nonlocal
variables (\ref{kdv1}--\ref{kdv2}). As explained in the previous section,
the
recursion operator is in effect a deformation of the equation structure.\\
As demonstrated there, this deformation is a form-valued vector field
and has to satisfy
\begin{align}\label{defor}
       \bar{\ell}_F^C(\CR) = 0.
\end{align}
In order to arrive at a nontrivial result as was explained for the KdV
equation, we have to introduce associated to the nonlocal variables 
$$P_0,~D P_0,~\oD P_0,~D \oD P_0,~Q_{1\over 2},~\ldots$$
their Cartan forms
\[
\om_{P_0},\ \om_{D P_0},\ \om_{\oD P_0},\ \om_{D \oD P_0},\ 
\om_{Q_{1\over 2}},\ \om_{D Q_{1\over 2}},\ \om_{\oD Q_{1\over 2}},\
\om_{D \oD Q_{1\over 2}},\]
\[ 
\om_{\oQ_{1\over 2}},\ \om_{D \oQ_{1\over 2}},\ \om_{\oD \oQ_{1\over
2}},\ \om_{D \oD\oQ_{1\over 2}},\ 
\om_{P_1},\ \om_{D P_1},\ \om_{\oD P_1},\ \om_{D \oD P_1}.\ 
\] 
These forms can straightforwardly be constructed using
eqs. (\ref{sec2:eq:kdv}) and (\ref{kdv0},\ref{kdv1}) in the line as
explained for the case of KdV-equation (see
e.g. eqs. (\ref{sec2:contforms}) and (\ref{sec3:nonlcontforms})) and
we shall present here  just a few of the associated explicit expresssions for
their differentiation rules\\
\begin{align*}
D\om_{J}&=-\om_{D J},\ D\om_{D J}=0,\ D\om_{\oD J}=-\om_{D\oD J},\\
\oD\om_{J}&=-\om_{\oD J},\ \oD\om_{D J}=-\om_{D\oD J}-\om_{J_1},\ \oD\oD\om_{J}=0,\\
D_x\om_{J}&=\om_{J_1},\ D_x\om_{D J}=\om_{D J_1},\ D_x\om_{\oD J}=\om_{\oD J_1},
\end{align*}
and similar for other Cartan forms, while additional signs arise  due to
{\bf odd} vector fields $D,\ \oD$.

Motivated by the results of the previous subsections and the grading
of the equation our search is for a one-form-valued vector field
whose defining function is of degree 3.\\
So besides the Cartan forms associated to the nonlocal variables
$P_0,\ldots$ which will account for the pseudo differential part of
the recursion operator, we also have to introduce the local Cartan forms
\[
\om_{J},\ \om_{D J},\ \om_{\oD J},\ \om_{D \oD J},\ 
\om_{J_{z}},\ \om_{D J_{z}},\ \om_{\oD J_{z}},\ \om_{D \oD J_{z}},\ 
\om_{J_{zz}},\ \om_{D J_{zz}},\ \om_{\oD J_{zz}},\ \om_{D \oD J_{zz}}\ 
\]
which will represent the pure differential part of the recursion
operator.\\
We have to remark that the coefficients of the one-form-valued vector
field are just functions dependent on all local and nonlocal
variables.\\
Moreover it should be noted that the representation of the vector field 
with respect to its form part is to be understood as to be equipped
with a right-module structure.
This will account for the correct action of contraction.\\
Since the coordinate  $P_0$ is of degree zero, this procedure requires
the introduction of approximately 2400, yet free, functions dependent on
this variable .\\
Now this quite extensive one-form-valued vector field has to satisfy 
eq. (\ref{defor}).
Although this is just the deformation condition which should be
satisfied, we decided, in order  to reduce the already extremely
extensive computations, first to  require  that the resulting
recursion operator performs its ``duties'', sending symmetries to
symmetries.\\
In order to achieve these goals we required the operator to satifies
the following requirements
\begin{align}
\CR(Y_1)=Y_3,\quad \CR(Y_3)=Y_5,\quad \CR(Y_5)=Y_7.
\end{align}
Due to these conditions it has been possible to fix all 2400 free functions
in the defining function $R$ of the one-form-valued vector field $\CR$.\\
 In
effect the condition $\CR(Y_5)=Y_7$ was itself sufficient to fix all
coefficients.\\
After this, the result has been substituted into (\ref{defor}), satisfying it
completely.\\
So, the final result is the following:\\
Starting form the defining function $R$ for the form-valued vector
field $\CR$, given by
\begin{align}
R=\sum_{\alpha} \om_{\alpha}\cdot \Phi_{\alpha}
\label{recop}
\end{align}
where $\alpha$ runs over $D \oD J_{zz},~\ldots, ~D J,~J,~P_0,~D
  P_0,~\ldots,D ~\oD P_1$, the coefficients are given by
\begin{align*}
\Phi_{D \oD P_1} &:= 0,\\
\Phi_{P_1} &:= J_{z},\\
\Phi_{\oD \oQ_{1\over 2}} &:= e^{(+)}\cdot D J\cdot D P_0 - D J\cdot
Q_{1\over 2},\\
\Phi_{\oD Q_{1\over 2}} &:=  - (\oD J\cdot D P_0 + D \oD J)\cdot
e^{(-)},\\
\Phi_{D \oQ_{1\over 2}} &:=  - (e^{(+)}\cdot D J\cdot \oD P_0 +
e^{(+)}\cdot D \oD J + e^{(+)}\cdot J_{z} + \oD J\cdot Q_{1\over 2}),\\
\Phi_{D Q_{1\over 2}} &:= e^{(-)}\cdot \oD J\cdot \oD P_0,\\
\Phi_{D \oD P_0} &:=  - \oD J\cdot D P_0 + 3\cdot J_{z},\\
\Phi_{P_0} &:=  - D J\cdot \oD J + 2\cdot J\cdot J_{z} + 2\cdot D \oD
J_{z} + J_{zz},\\
\Phi_{\oD P_1} &:=  - D J,\\
\Phi_{D P_1} &:=  - \oD J,\\
\Phi_{D \oD \oQ_{1\over 2}} &:=  - e^{(+)}\cdot D J/2,\\
\Phi_{D \oD Q_{1\over 2}} &:=  - e^{(-)}\cdot \oD J/2,\\
\Phi_{\oQ_{1\over 2}} &:= 2\cdot e^{(+)}\cdot D P_0\cdot D \oD J + 2\cdot
e^{(+)}\cdot D P_0\cdot J_{z} + e^{(+)}\cdot D J_{z} - 2\cdot
e^{(+)}\cdot D J\cdot D P_0\cdot \oD P_0 \\
&\ \ \ \ \ - e^{(+)}\cdot D J\cdot D \oD P_0 - Q_{1\over 2}\cdot J_{z} -
\oD J\cdot D Q_{1\over 2} - D J\cdot \oD Q_{1\over 2},\\
\Phi_{Q_{1\over 2}} &:= e^{(-)}\cdot ( - 2\cdot \oD P_0\cdot D \oD J -
\oD J_{z} - 2\cdot \oD J\cdot D P_0\cdot \oD P_0 + \oD J\cdot J + \oD
J\cdot D \oD P_0),\\
\Phi_{\oD P_0}  &:=  - D P_0\cdot J_{z} - D J_{z} - D J\cdot J + D J\cdot
D \oD P_0,\\
\Phi_{D P_0} &:= \oD J_{z} - 2\cdot \oD J\cdot J,\\
\Phi_{J_{zz}} &:= 1,\\
\Phi_{D \oD J_{z}} &:= 0,\\
\Phi_{J_{z}} &:= 2\cdot J,\\
\Phi_{D \oD J} &:= 4\cdot J,\\
\Phi_{J} &:= J^2 + 3\cdot D \oD J + 3\cdot J_{z},\\
\Phi_{\oD J_{z}}  &:= 0,\\
\Phi_{D J_{z}} &:= 0,\\
\Phi_{\oD J} &:= 0,\\
\Phi_{D J} &:= ( - \oD J)/2.\\
\end{align*}
This now finishes the longstanding problem of the existence of the
recursion operator for the $N=2$ supersymmetric $\alpha=1$ KdV-equation.\\
Transformation of the formvaluedness to the Fr\'echet derivatives
leads to the presentation of the recursion operator in classical form
(\cite{kersor}).
We have checked, as mentioned before, that the form-valued vector
field (\ref{recop}) satisfies eq. (\ref{defor}) and so indeed gives the
proper recursion operator for symmetries of the $N=2$ ${\alpha}=1$
KdV-hierarchy.

\subsection{Factorization of the Recursion Operator and the Bi-Hamiltonian
structure}
Here we shall present the factorization of the recursion operator
obtained in last section.\\
Factorization in this respect means
\begin{align}
R=J_2\cdot J_0^{-1}
\end{align}
where $J_2$ is the second Hamiltonian structure \cite{lm}
\begin{eqnarray}
J_2\equiv \frac{1}{2}[D, \overline D ~]\partial +\overline DJD 
+DJ\overline D + \partial J +J\partial
\label{hamstr2}
\end{eqnarray}
and $J_0$ will be the zero Hamiltonian structure.\\
We assume $J_0^{-1}$ to be a one-form-valued function, i.e., in effect 
a pseudo--differential operator, the pseudo--part of which is realized
through the Frechet derivatives of the nonlocal variables.\\
So we describe $J_0^{-1}$ in a similar way as the defining function of 
the deformation structure 
\begin{align}
J_0^{-1}=\sum_{\alpha} \om_{\alpha} \cdot \Phi^0_{\alpha}
\end{align}
where $\alpha$ runs over local and nonlocal variables and
$\Phi^0_{\alpha}$ 
being function of appropriate degree.\\
A rather straightforward computation does lead to the following result:\\
$\alpha$  and  the associated $\Phi^0_{\alpha}$ are given by\\
\begin{align*}
\Phi^0_{P_1} &:= 1,\\
\Phi^0_{\oD Q_{1\over 2}} &:=  - e^{(-)})/2,\\
\Phi^0_{D \oQ_{1\over 2}} &:= e^{(+)}/2,\\
\Phi^0_{D \oD P_0} &:= 3,\\
\Phi^0_{P_0} &:= J,\\
\Phi^0_{\oQ_{1\over 2}} &:=  - (e^{(+)}\cdot D P_0 + Q_{1\over 2}),\\
\Phi^0_{Q_{1\over 2}} &:=  - e^{(-)}\cdot\oD P_0,\\
\Phi^0_{\oD P_0} &:=  - D P_0,\\
\Phi^0_{J} &:= 3/2.
\end{align*}
If we use now Frechet derivatives associated to the occurring nonlocal 
variables in the presentation, we arrive at \cite{kersor} \\

\begin{eqnarray}
J^{-1}_0 = 
[D,{\overline D}~]{\partial}^{-1} + {\partial}^{-1} J_2 {\partial}^{-1}
+\frac{1}{2}{\overline{f}_{\frac{1}{2}}}^{T}{\partial}^{-1}f_{\frac{1}{2}}
-\frac{1}{2}{f^{T}_{\frac{1}{2}}}{\partial}^{-1}\overline{f}_{\frac{1}{2}}
\label{firstHSTR}
\end{eqnarray}
where $f_{\frac{1}{2}},\overline{f}_{\frac{1}{2}}$ are the Fr\'echet
derivatives of $Q_{1\over 2}, \oQ_{1\over 2}$ respectively.\\
The factorization of the recursion operator for symmetries  has wider
applicability in the construction of Hamiltonian operators and  will be discussed more deeply  elsewhere.
\section{Conclusion}
We gave an outline of the theory of deformations of the equation structure of
differential equations, leading to the construction of recursion operators
for symmetries of such equations. The extension of this theory to the
nonlocal setting of differential equations is essential for getting
nontrivial results. The theory has been applied to the construction of the
recursion operator for symmetries for a coupled KdV--mKdV system, leading to
a highly nonlocal result for this system. Moreover the appearance of
nonpolynomial nonlocal terms in all results, e.g., conservation laws,
symmetries and recursion operator is striking and reveals some unknown and
intriguing underlying structure of the equations.

{}~

{}~

\noindent{\bf Acknowledgments.}
A.S. is grateful to University of Twente for the hospitality extended to
him during this research. This work was partially supported by the grants
NWO NB 61-491, FOM MF 00/39, RFBR 99-02-18417, RFBR-DFG 02-02-04002,
RFBR-CNRS 98-02-22034, PICS Project No. 593, Nato Grant No. PST.CLG 974874
and the Heisenberg-Landau program.                       

{}~

\newpage

\section{Appendix: The Second layer of Nonlocalities}
Here we present the second set of sixteen nonlocal variables,
conservation laws and Hamiltonians.\\
We shall present here the results for\\
\begin{align*}
&P_2,~D P_2,~\oD P_2,~D \oD P_2;\\
&Q_{5\over 2},~D Q_{5\over 2},~\oD Q_{5\over 2},~D \oD Q_{5\over 2};\\
&\oQ_{5\over 2},~D \oQ_{5\over 2},~\oD \oQ_{5\over 2},
~D \oD \oQ_{5\over 2};\\
&P_3,~D P_3,~\oD P_3,~D \oD P_3.
\end{align*}

The explicit formulae for $(P_2)_z,~(Q_{5\over 2})_z,~(\oQ_{5\over 2})_z$
are 

$(P_2)_z := ((-2)\cdot ( - e^{2(+)}\cdot D J\cdot \oQ_{1\over 2}\cdot J -
e^{2(+)}\cdot D J\cdot \oQ_{1\over 2}\cdot D \oD P_0 - e^{2(+)}\cdot D
J\cdot \oD P_0\cdot D \oQ_{1\over 2} - 2\cdot e^{2(+)}\cdot D J\cdot D
P_0\cdot \oD \oQ_{1\over 2} + e^{(+)}\c
dot \oD J\cdot Q_{1\over 2}\cdot D \oQ_{1\over 2} + e^{(+)}\cdot \oD
J\cdot D P_0\cdot J + e^{(+)}\cdot \oD J\cdot D P_0\cdot D \oD P_0 +
e^{(+)}\cdot D J\cdot \oQ_{1\over 2}\cdot \oD Q_{1\over 2} - e^{(+)}\cdot
D J\cdot \oD P_0\cdot D \oD P_0 - 2\cdot e^
{(+)}\cdot D J\cdot \oD J + \oD J\cdot Q_{1\over 2}\cdot D \oD P_0 - \oD
J\cdot D P_0\cdot \oD Q_{1\over 2}))/e^{(+)},$\\

$(Q_{5\over 2})_z := (2\cdot ( - e^{2(+)}\cdot D P_1\cdot D \oD J -
e^{2(+)}\cdot D
P_1\cdot J_{z} + e^{2(+)}\cdot D P_0\cdot D \oD J\cdot D \oD P_0 +
e^{2(+)}\cdot D P_0\cdot J_{z}\cdot D
\oD P_0 - 2\cdot e^{2(+)}\cdot D J\cdot \oD P_0\cdot D P_1 -
e^{2(+)}\cdot D J\cdot J\cdot \cdot 2 -
e^{2(+)}\cdot D J\cdot J\cdot D \oD P_0 - 2\cdot e^{2(+)}\cdot D J\cdot D
\oD J - 2\cdot e^{2(+)}\cdot D J\cdot J_{z} + e^{(+)}\cdot Q_{1\over 2}\cdot J\cdot D \oD J + e^{(+)}\cdot Q_{1\over 2}\cdot D \oD J\cdot D \oD P_0 - e^{(+)}\cdot \oD P_0\cdot D \oD
J\cdot D Q_{1\over 2} + 2\cdot e^{(+)}\cdot \oD J\cdot D P_0\cdot
Q_{1\over 2}\cdot J + 2\cdot e^{(+)}\cdot \oD J\cdot D P_0\cdot Q_{1\over
2}\cdot D \oD P_0 - e^{(+)}\cdot \oD J\cdot J\cdot D Q_{1\over 2} -
2\cdot e^{(+)}\cdot \oD J\cdot D \oD P_0\cdot 
D Q_{1\over 2} - 2\cdot e^{(+)}\cdot D J\cdot \oD P_0\cdot Q_{1\over
2}\cdot D \oD P_0 - 2\cdot e^{(+)}\cdot D J\cdot \oD J\cdot Q_{1\over 2}
- 2\cdot \oD J\cdot \oD P_0\cdot Q_{1\over 2}\cdot D Q_{1\over
2}))/e^{(+),}$\\

$(\oQ_{5\over 2})_z := (2\cdot ( - e^{(+)}\cdot \oQ_{1\over 2}\cdot D \oD
J\cdot D \oD P_0
- e^{(+)}\cdot \oQ_{1\over 2}\cdot J_{z}\cdot D \oD P_0 + 2\cdot
e^{(+)}\cdot D P_0\cdot \oD
P_0\cdot \oQ_{1\over 2}\cdot J_{z} + 2\cdot e^{(+)}\cdot \oD J\cdot D
P_0\cdot \oQ_{1\over 2}\cdot D \oD
P_0 + 2\cdot e^{(+)}\cdot D J\cdot \oD P_0\cdot \oQ_{1\over 2}\cdot J -
2\cdot e^{(+)}\cdot D J\cdot \oD
P_0\cdot \oQ_{1\over 2}\cdot D \oD P_0 - 2\cdot e^{(+)}\cdot D J\cdot \oD
J\cdot \oQ_{1\over 2} -
e^{(+)} \cdot D J\cdot D \oD P_0\cdot \oD \oQ_{1\over 2} - \oD P_1\cdot D
\oD J - \oQ_{1\over 2}\cdot D \oD J\cdot \oD Q_{1\over 2} + Q_{1\over
2}\cdot D \oD J\cdot \oD \oQ_{1\over 2} - 2\cdot \oD J_{z}\cdot D \oD P_0
- 2\cdot \oD J\cdot \oD P_0\cdot \oQ_
{1\over 2}\cdot D Q_{1\over 2} - 2\cdot \oD J\cdot D P_0\cdot \oD P_1 -
2\cdot \oD J\cdot D P_0\cdot \oQ_{1\over 2}\cdot \oD Q_{1\over 2} +
2\cdot \oD J\cdot D P_0\cdot Q_{1\over 2}\cdot \oD \oQ_{1\over 2} +
2\cdot \oD J\cdot D P_0\cdot \oD P_0\cdot J + 3
\cdot \oD J\cdot J\cdot D \oD P_0 - \oD J\cdot D Q_{1\over 2}\cdot \oD
\oQ_{1\over 2}))/e^{(+).}$\\
Other quantities can be obtained by action of $D$ and $\oD$.\\

The nonlocal variable $P_3$ results from the Poisson bracket of other
Hamiltonians, i.e.,\\
\begin{align*}
&\{Q_{1\over 2},\oQ_{5\over 2}\}.
\end{align*}
The general Poisson algebra structure of the Hamiltonians will be
discussed elsewhere \cite{kersor2}.\\


\end{document}